\documentclass[preprint,showpacs,preprintnumbers,amsmath,amssymb]{revtex4}
\usepackage{graphicx}
\usepackage{dcolumn}
\usepackage{bm}
\usepackage{amssymb}
\usepackage{amsmath}
\usepackage{latexsym}

\begin{document}

\title{On the Darwin instability effect in binary systems}

\author{V.V. Sargsyan $^{1,2}$, H. Lenske$^{2}$, G.G. Adamian$^{1}$, and N.V. Antonenko$^{1}$}
\affiliation{
$^{1}$Joint Institute for Nuclear Research, 141980 Dubna, Russia,\\
$^{2}$Institut f\"ur Theoretische Physik der
Justus--Liebig--Universit\"at, D--35392 Giessen, Germany}

\date{\today}

\begin{abstract}
The Darwin instability effect in the binary systems (planets, stars, and galaxies)
is analyzed within the model based on the Regge-like laws.
New analytical formulas  are presented for the relative distance between  components of the binary
and orbital rotation period of the binary.
\end{abstract}


\maketitle


It is now commonly believed that the contact binaries, for example, the W UMa binaries,
end their evolution by merging into a single star \cite{Cher:2013}.
The dissipation of the orbital energy in the initial  violent
phase of  merger resulted in the  luminous red nova V1309 Sco  observed in 2008.
The luminous
red novae have recently been identified as a distinct class of
stellar transients \cite{Kulkarni:2007}. They are characterized by
relatively long outbursts with spectral distributions centered in the
red, ranging in luminosities  between classical novae and
supernovae.
The observations in the Optical Gravitational Lensing Experiment
have revealed a shape of the light curve characteristic for  contact
binary system with an exponentially decreasing orbital period
\cite{Tylenda:2011} and confirmed   the earlier
conjecture of Ref. \cite{Soker:2003} that the luminous red
novae  arise from merging contact binary stars.
The spectacular  case is KIC 9832227 which
 was predicted \cite{Molnar:2017} to be   merge in 2022,
 enlightening the sky as a red nova. The
 compact  binaries composed of white dwarfs, neutron stars, and
black holes  eventually merge through
gravitational wave emission \cite{Abbott:2016a,Abbott:2016b,Abbott:2017}.

The details
 of   specific
mechanism, which triggers the merger of the contact
binary components,  are still controversial \cite{Sana:2012,Kocha:2014,Fabrycky:2007,Tokovinin:2006,Eggleton:2012,Pejcha:2017}.
There is assumption
\cite{Eggleton:2012} on the gradual mass transfer from  less
massive (and hence smaller radius) secondary star to the primary one (heavier star).
This mass transfer is driven by the structural
change in the secondary star caused by the energy   received from the
primary star. The contact is sustained by the  magnetic
braking or thermonuclear evolution.
When the mass ratio is extreme enough for the Darwin instability,
 a merger starts that triggers
  the outburst in red novae
\cite{Darwin:1879,Tylenda:2011}.
The Darwin instability happens when the spin angular
 momentum of the system is more than one third of the orbital angular momentum.
This instability plays a role
once the mass ratio becames small enough that the companion star can
no longer keep the primary star synchronously rotating via the tidal
interaction.
For most of the primary massive stars,
this occurs at the mass ratio $q=M_2/M_1< 0.1$ \cite{Rasio:1995}.
There is   alternative scenario \cite{Stepien:2011}: At
first contact in a binary system, a brief, but intense, mass transfer
sets in  changing  originally more massive star into   less massive
one. This process may oscillate until eventually a stable contact
configuration is reached.
Also the dynamic mass transfer without the Darwin
instability \cite{DSouza:2006} and mergers triggered by a tidal runaway
based on a non-equilibrium response to tidal dissipation
\cite{Koenigsberger:2016} have been investigated.
Thus, there is a large
quest for  detailed   observational and theoretical
investigations \cite{Stepien:2011,DSouza:2006,Koenigsberger:2016}.

As shown in Refs. \cite{Murad1,Murad2,Murad3},   the angular momentum $J$ of many astronomical objects, from planets to
clusters of galaxies, and, possibly, the universe as a whole, can be predicted by simple Regge-like laws
from the mass $M$ of the object. The Regge-theory proved to be very influential in the development of elementary-particle physics.
In Refs. \cite{Murad1,Murad2,Murad3}, the cosmic analog of the Chew-Frautschi plot with two important cosmological
Eddington and Chandrasekhar points on it has been constructed.
The application of Regge ideas to astrophysics
  has   shown that the spins of   planets and stars are well described
by the Regge-like law for a sphere ($J\sim M^{4/3}$), while the spins of galaxies 
and clusters of galaxies obey the Regge-like law for a disk ($J\sim M^{3/2}$)
\cite{Murad1,Murad2,Murad3}.
 In contrast to earlier semi-phenomenological approaches
these expressions contain only fundamental constants as the parameters and are independent of any fitted  empirical
quantities.
The aim of the present article is to study the Darwin instability effect in the binary star or galaxy by using
the model of Refs. \cite{Murad1,Murad2,Murad3} based on the Regge-theory.


The total angular momentum  ${\bf J}_{\rm tot}$  of   binary system is the sum of   
orbital angular momentum ${\bf L}$ and the spins ${\bf S}_k$ ($k=$1,2) of the individual components:
\begin{eqnarray}
{\bf J}_{\rm tot} = {\bf L} + {\bf S}_1 + {\bf S}_2.
\label{eq_1}
\end{eqnarray}
The $J_{\rm tot}$ and $S_k$ are expressed using the Regge-like law  for stars and planets ($n=3$) or galaxies ($n=2$):
\begin{eqnarray}
J_{\rm tot}=\hbar\left(\frac{M}{m_p}\right)^{(1+n)/n}
\label{eq_2}
\end{eqnarray}
and
\begin{eqnarray}
S_k=\hbar\left(\frac{M_k}{m_p}\right)^{(1+n)/n},
\label{eq_3}
\end{eqnarray}
where  $\hbar$, $m_p$, $M_k$ ($k=$1,2),   and $M=M_1+M_2$ are the  Planck constant,
masses of  proton and astrophysical objects (planets, stars or galaxies), and the total mass of system, respectively.
The maximum (the antiparallel orbital and spins angular momenta) and minimum (the  parallel orbital and spins angular momenta)
 orbital angular momenta are
\begin{eqnarray}
L_{\rm max}=J_{\rm tot}+S_1+S_2
\label{eq_4}
\end{eqnarray}
and
\begin{eqnarray}
L_{\rm min}=J_{\rm tot}-S_1-S_2,
\label{eq_5}
\end{eqnarray}
respectively.
Using the mass asymmetry (mass transfer) coordinate $\eta=(M_1-M_2)/M$
instead of masses $M_1=\frac{M}{2}(1+\eta)$ and $M_2=\frac{M}{2}(1-\eta)$ \cite{IJMPE} and Eqs. (\ref{eq_3})--(\ref{eq_5}), we derive
\begin{eqnarray}
\frac{S_1+S_2}{L_{\rm min}}=\frac{(1+\eta)^{(1+n)/n}+(1-\eta)^{(1+n)/n}}{2^{(1+n)/n}-(1+\eta)^{(1+n)/n}-(1-\eta)^{(1+n)/n}},\\
\frac{S_1+S_2}{L_{\rm max}}=\frac{(1+\eta)^{(1+n)/n}+(1-\eta)^{(1+n)/n}}{2^{(1+n)/n}+(1+\eta)^{(1+n)/n}+(1-\eta)^{(1+n)/n}}.\\
\label{eq_6}
\end{eqnarray}
At $\eta=0$, we have
$$\frac{S_1+S_2}{L_{\rm min}}=\frac{1}{2^{1/n}-1}>1$$ and
$$\frac{S_1+S_2}{L_{\rm max}}=\frac{1}{2^{1/n}+1}>\frac{1}{3}.$$
For the symmetric binary star (planet) and binary galaxy, $(S_1+S_2)/L_{\rm max}\approx$ 0.44 and 0.41, respectively.
At $\eta=1$, we have
$$\frac{S_1+S_2}{L_{\rm min}}\to\infty$$ and
$$\frac{S_1+S_2}{L_{\rm max}}=\frac{1}{2}.$$
As follows from last two expressions,  for very asymmetric binaries, the ratios $(S_1+S_2)/L_{\rm max,min}$ almost independent of the value of $n$.
According to Ref.  \cite{Rasio:1995}, the Darwin instability can occur  when the binary mass ratio is very small  ($q=M_2/M_1< 0.1$) or the
mass asymmetry  is very large ($\eta=(1-q)/(1+q)>0.82$).
As seen in Fig. 1, the ratios $(S_1+S_2)/L_{\rm max}$ and $(S_1+S_2)/L_{\rm min}$
 continuously increases  with $\eta$ from 0 to 1.
Because their absolute values are larger than 1/3,  all possible binary stars (planets) or binary galaxies, independently of their mass asymmetry $\eta$,
should have the Darwin instability ($S_1 + S_2 \ge \frac{1}{3}L$) and, correspondingly, should merge.
However,  the observations do not support this conclusion
which probably  means that there is no the Darwin instability effect in such binary systems and, correspondingly, the mechanism of merger
has other origin.

Note that in the cases of    antiparallel spins with $L_1=J_{tot}+S_1-S_2$   and
$L_2=J_{tot}-S_1+S_2$ (Fig. 1), the ratios $|S_2-S_1|/L_{1}$ and $|S_1-S_2|/L_{2}$
are larger than $\frac{1}{3}$ for the asymmetric binaries with $|\eta|\ge 0.5$.

As seen in Fig. 2, the dependencies of $L_{\rm max}$, $L_{\rm min}$, $L_{1}$, and $L_{2}$ on mass asymmetry have different behavior.
The evolution of system in  mass asymmetry (mass transfer) can increase or decrease the orbital angular momentum.
For example, at $\eta\to 0$  the binary system has  smaller $L=L_{\rm max}$.
The observations of the dependence of $L$ on $\eta$ may be useful to distinguish the difference between the orientations
of orbital and spins angular momenta.

Employing Eqs. (\ref{eq_1})--(\ref{eq_3}) and results of Refs. \cite{nash}, we obtain new analytical formulas
for the relative distance between the components of the binary
\begin{eqnarray}
R_m&=&\frac{ML^2}{GM_1^2M_2^2}\nonumber\\
&=&\frac{\hbar^2M}{GM_1^2M_2^2}
\left[\left(\frac{M}{m_p}\right)^{(1+n)/n}+\epsilon_1\left(\frac{M_1}{m_p}\right)^{(1+n)/n}+\epsilon_2\left(\frac{M_2}{m_p}\right)^{(1+n)/n}\right]^2,\nonumber\\
&=&\frac{2\hbar^2}{Gm_p^3}\left(\frac{M}{2m_p}\right)^{(2-n)/n}\frac{[2^{(1+n)/n}+\epsilon_1 (1+\eta)^{(1+n)/n}+\epsilon_2(1-\eta)^{(1+n)/n}]^2}{[1-\eta^2]^2}
\label{eq_7}
\end{eqnarray}
at $R_m>R_t=R_1+R_2$ ($R_k$ are the radii of binary components)
and
\begin{eqnarray}
R_m&=&\left(\frac{\hbar M g^3}{GM_1^2M_2^2}\right)^{1/4}
\left[M_1^{l} + M_2^{l}\right]^{3/4}
\left[\left(\frac{M}{m_p}\right)^{(1+n)/n}+\epsilon_1\left(\frac{M_1}{m_p}\right)^{(1+n)/n}+\epsilon_2\left(\frac{M_2}{m_p}\right)^{(1+n)/n}\right]^{1/2},\nonumber\\
&=&\left(\frac{2\hbar m_p^{3l-3}g^3}{G}\right)^{1/4}\left(\frac{M}{2m_p}\right)^{(2+[3l-1]n)/n}\left[(1+\eta)^l+(1-\eta)^l\right]^{3/4}\nonumber\\
&\times&\frac{[2^{(1+n)/n}+\epsilon_1 (1+\eta)^{(1+n)/n}+\epsilon_2 (1-\eta)^{(1+n)/n}]^{1/2}}{[1-\eta^2]^{1/2}}
\label{eq_8}
\end{eqnarray}
at $R_m\le R_t$,
and for the orbital rotation period in the binary
\begin{eqnarray}
P_{\rm orb}&=& 2\pi\left(\frac{ R_m^3}{G M }\right)^{1/2}   \nonumber\\
&=&\frac{2\pi\hbar^3M}{G^2M_1^3M_2^3}
\left[\left(\frac{M}{m_p}\right)^{(1+n)/n}+\epsilon_1\left(\frac{M_1}{m_p}\right)^{(1+n)/n}+\epsilon_2\left(\frac{M_2}{m_p}\right)^{(1+n)/n}\right]^3\nonumber\\
&=&\frac{4\pi\hbar^3}{G^2m_p^5}\left(\frac{M}{2m_p}\right)^{(3-2n)/n}\frac{[2^{(1+n)/n}+\epsilon_1 (1+\eta)^{(1+n)/n}+\epsilon_2 (1-\eta)^{(1+n)/n}]^3}{[1-\eta^2]^3}
\label{eq_9}
\end{eqnarray}
at $R_m>R_t$
and
\begin{eqnarray}
P_{\rm orb}&=& 2\pi\left(\frac{ R_t^3}{G M }\right)^{1/2}  \nonumber\\
&=&2\pi\left(\frac{g^3\left[M_1^{l} + M_2^{l}\right]^{3}}{GM}\right)^{1/2}\nonumber\\
&=&2\pi\left(\frac{g^3 M^{3l-1}}{2^{3l}G}\right)^{1/2}\left[(1+\eta)^l+(1-\eta)^l\right]^{3/2}\nonumber\\
&=&
2\pi\left(\frac{g^3 m_p^{3l-1}}{2G}\right)^{1/2}\left(\frac{M}{2m_p}\right)^{(3l-1)/2}\left[(1+\eta)^l+(1-\eta)^l\right]^{3/2}
\label{eq_10}
\end{eqnarray}
at $R_m<R_t$ \cite{nash}.
Here,  $G$, $R_t=R_1+R_2$, and $R_k$ ($k=1,2$) are the gravitational constant, touching distance, and radius of the component of binary, respectively.
The values $\epsilon_1=\epsilon_2=1$ and $\epsilon_1=\epsilon_2=-1$ correspond to the cases of antiparallel and parallel orbital and spins angular momenta.
The values $\epsilon_1=-\epsilon_2=1$ and $\epsilon_1=-\epsilon_2=-1$ correspond to the cases of antiparallel  spins.
The observational data  result in the relationship
$$R_k=gM_k^{l}$$
between the radius and mass of the star, where the constants $l=\frac{2}{3}$
and $g=R_{\odot}/M_\odot^{2/3}$ ($M_{\odot}$ and $R_{\odot}$ are   mass  and radius of the Sun) \cite{Vasil:2012}  and the galaxy,
where the constant $l$ depending on mass is in the interval  $\left[\frac{2}{5},\frac{2}{3}\right]$ \cite{Karachentsev}.
As seen in Figs. 3--5, at   $R_m>R_t$ ($R_m\le R_t$),
 $R_m$ decreases ($R_t$ increases) with decreasing $|\eta|$
and, finally, the $P_{orb}$ decreases (increases).
At   $R_m>R_t$, the dependence of $P_{orb}$ as a function of mass asymmetry has
similar behavior in the cases when the orbital and spin
angular momenta are antiparallel and parallel.
At $R_m\le R_t$, the value of $P_{orb}$ does not depend on orientations of  orbital and spin
angular momenta.

\begin{figure} [ht]
\centering
{\includegraphics[width=0.49\linewidth]{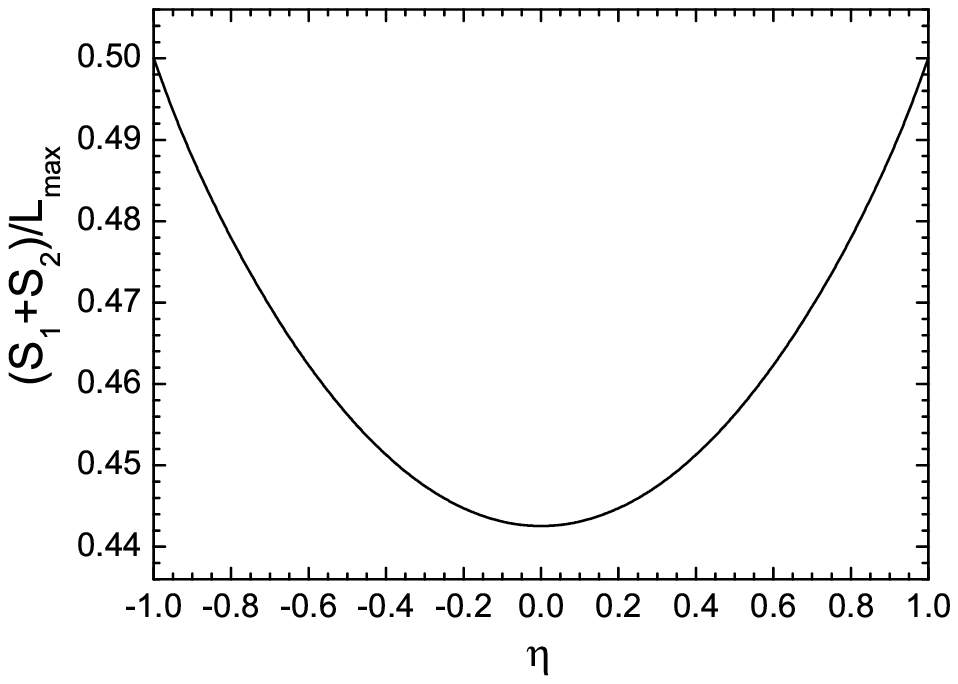}
\includegraphics[width=0.49\linewidth]{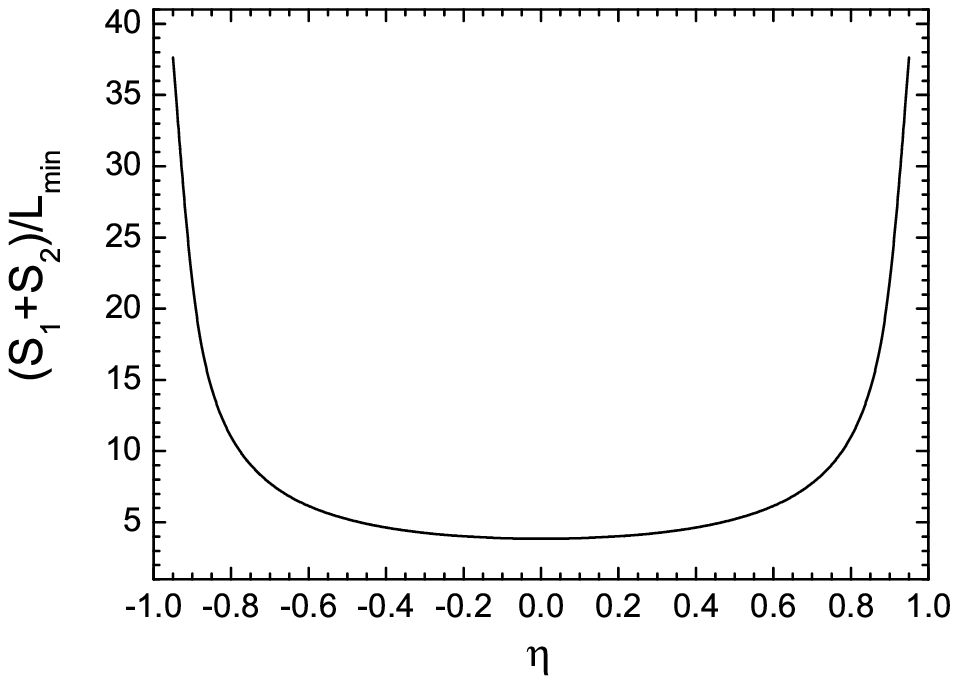}}
\includegraphics[width=0.49\linewidth]{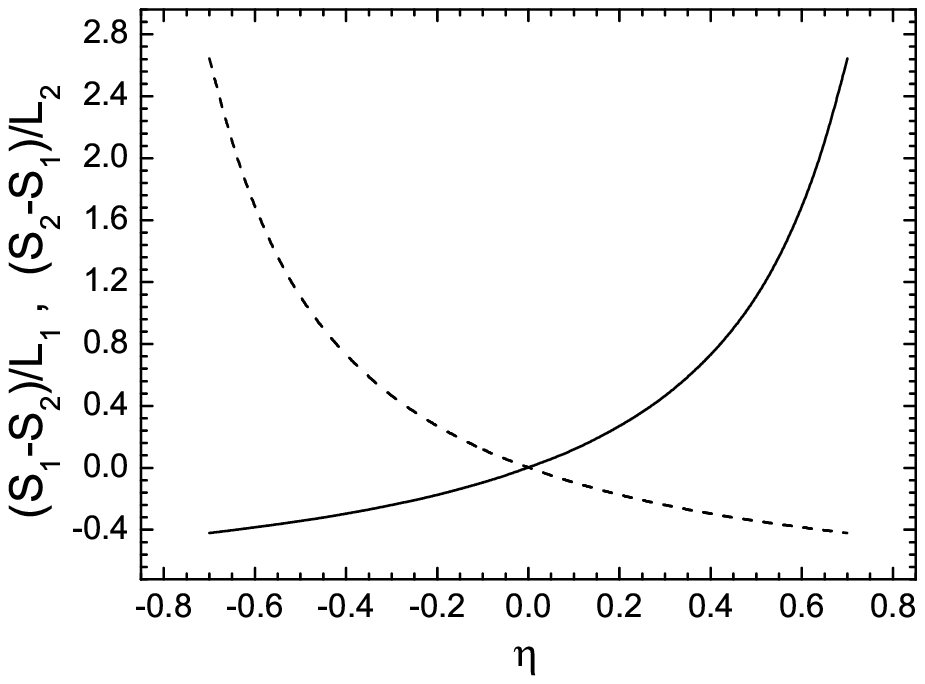}
\caption{The calculated ratios $(S_1+S_2)/L_{\rm max}$, $(S_1+S_2)/L_{\rm min}$, $(S_1-S_2)/L_{2}$ (solid line),
and  $(S_2-S_1)/L_{1}$  (dashed line) as   functions of mass asymmetry. In the cases of antiparallel spins, $L_1=J_{tot}+S_1-S_2$ and
$L_2=J_{tot}-S_1+S_2$.
}
\label{1_fig}
\end{figure}
\begin{figure} [ht]
\centering
{\includegraphics[width=0.49\linewidth]{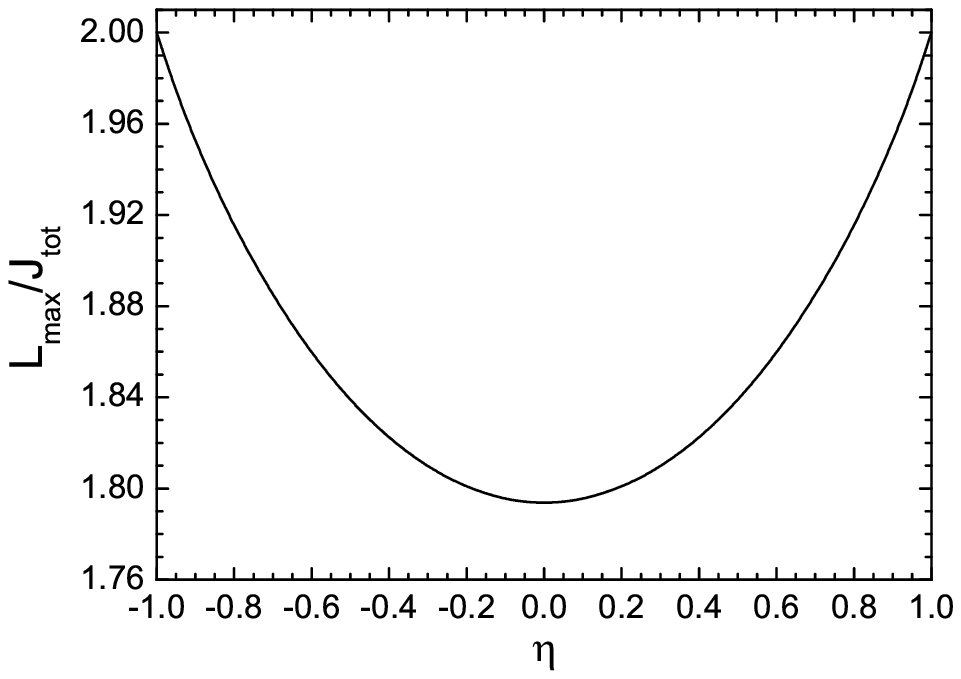}
\includegraphics[width=0.49\linewidth]{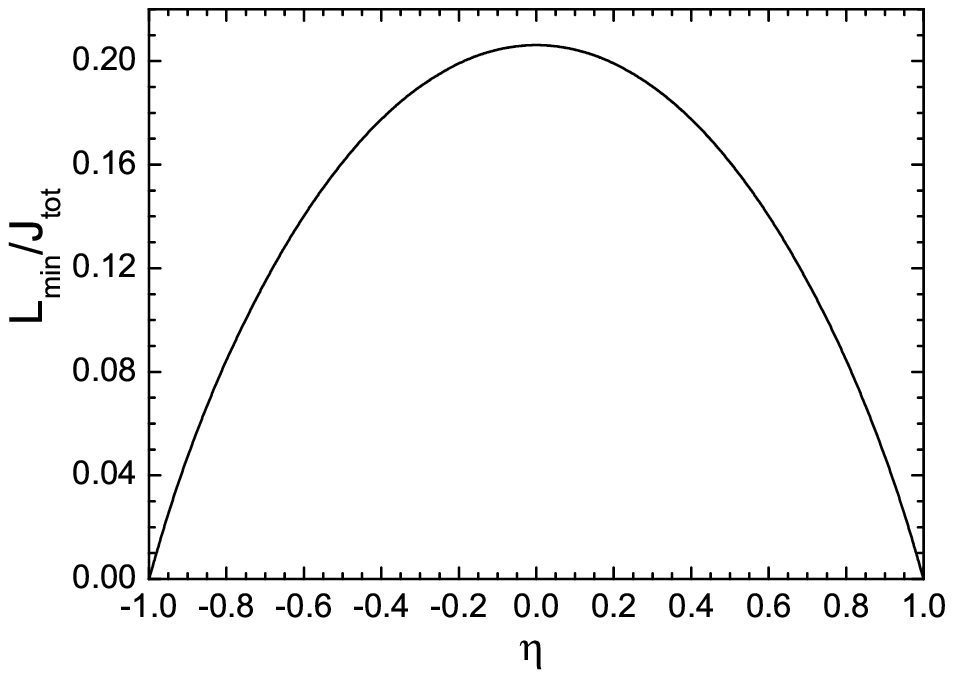}}
\includegraphics[width=0.49\linewidth]{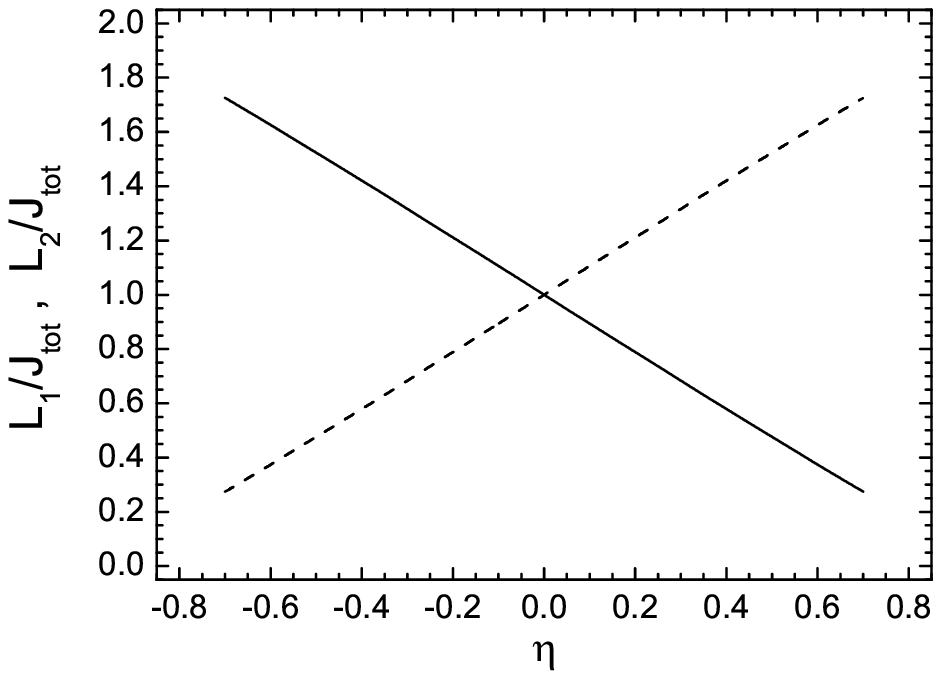}
\caption{The calculated ratios $ L_{\rm max}/J_{\rm tot}$, $ L_{\rm min}/J_{\rm tot}$,
$L_{2}/J_{\rm tot}$ (solid line),
and   $L_{1}/J_{\rm tot}$ (dashed line)   as   functions of mass asymmetry.
}
\label{2_fig}
\end{figure}
\begin{figure} [ht]
\centering
{\includegraphics[width=0.49\linewidth]{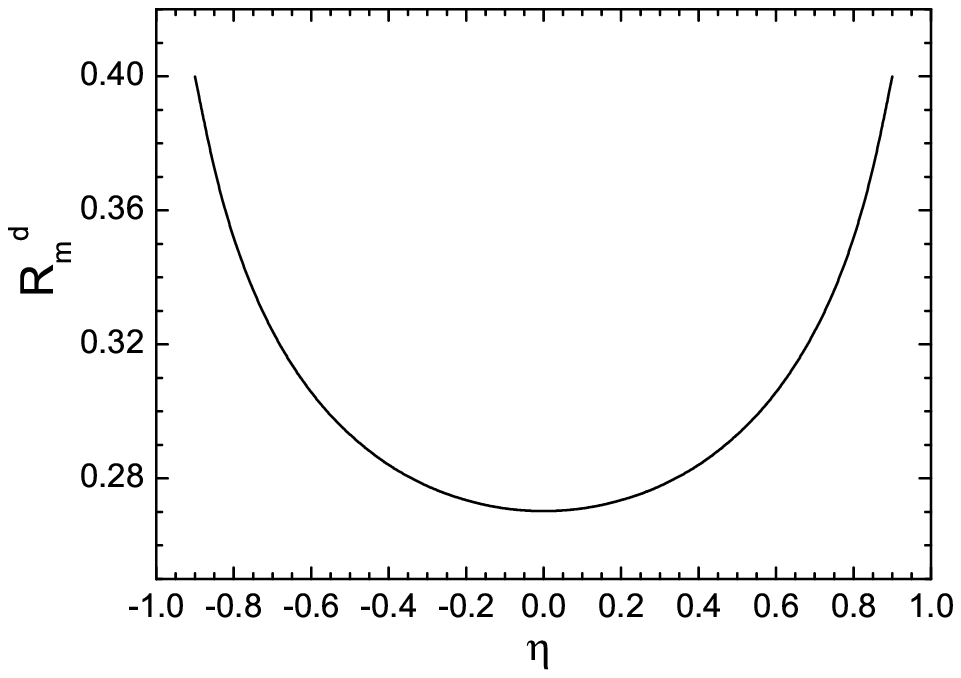}
\includegraphics[width=0.49\linewidth]{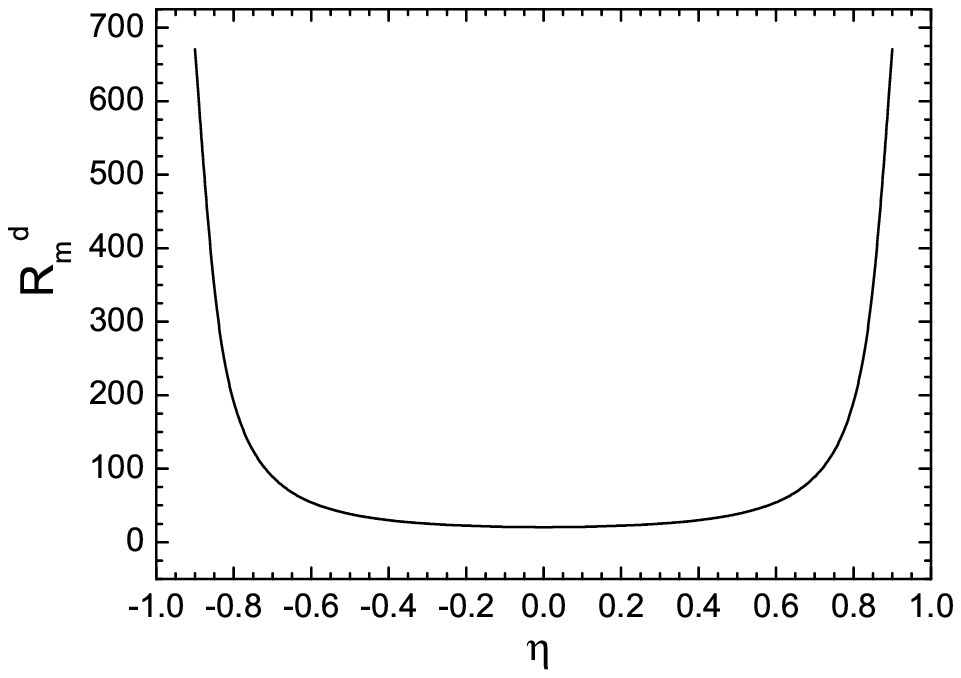}}
{\includegraphics[width=0.49\linewidth]{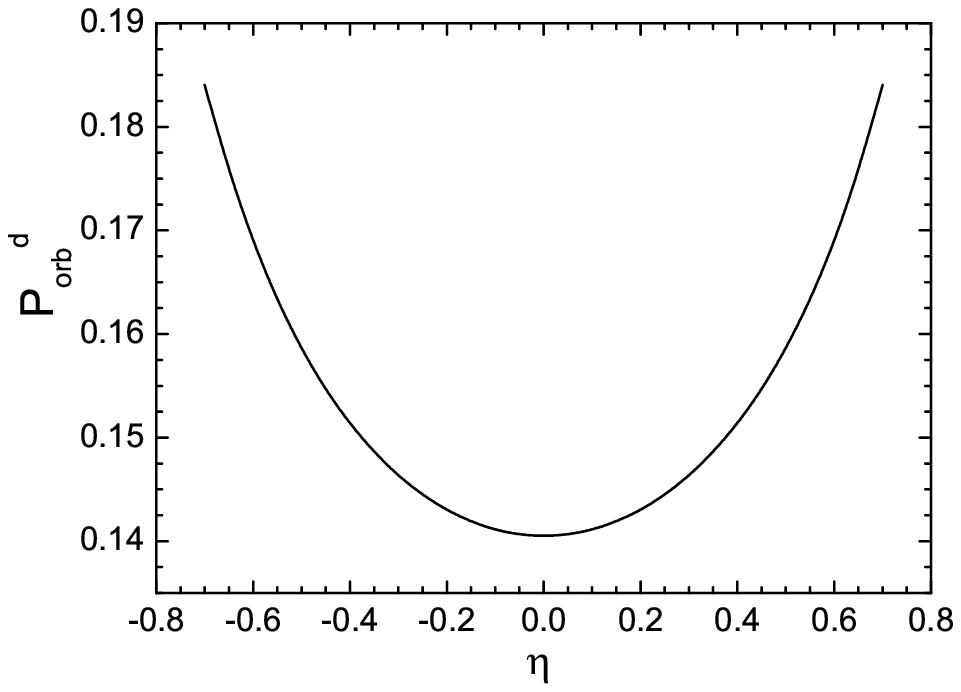}
\includegraphics[width=0.49\linewidth]{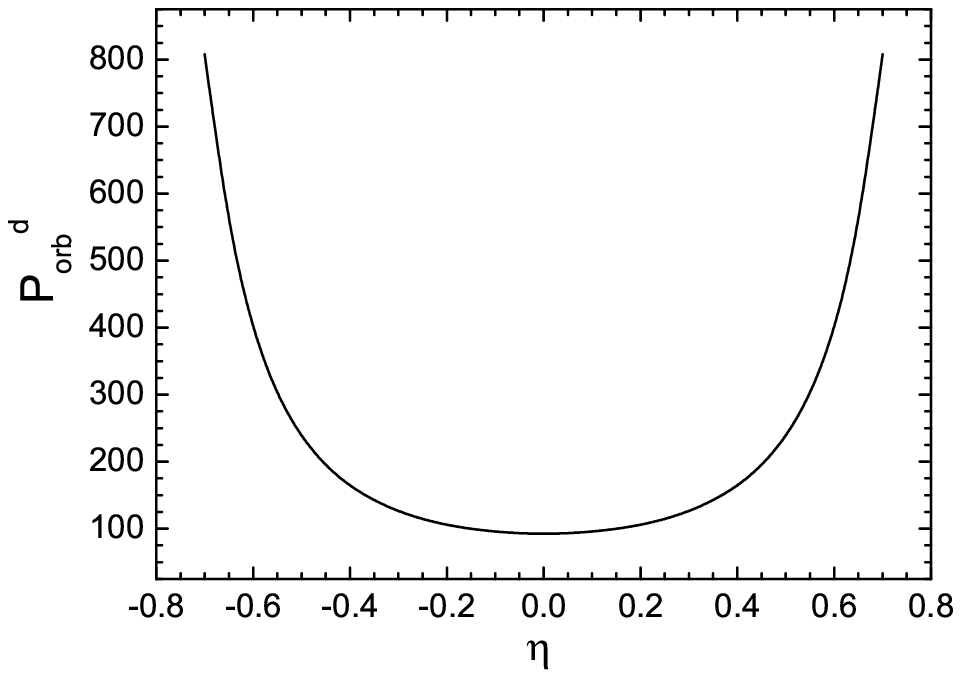}}
\caption{The calculated dimensionless relative distances $R_{m}^{d}$
and orbital rotation periods
$ P_{orb}^{d}$ as   functions of mass asymmetry at $R_m>R_t=R_1+R_2$.
The left and right sides correspond to the systems
with the parallel and antiparallel, respectively, orbital and spins
angular momenta.
}
\label{3_fig}
\end{figure}
\begin{figure} [ht]
\centering
{\includegraphics[width=0.49\linewidth]{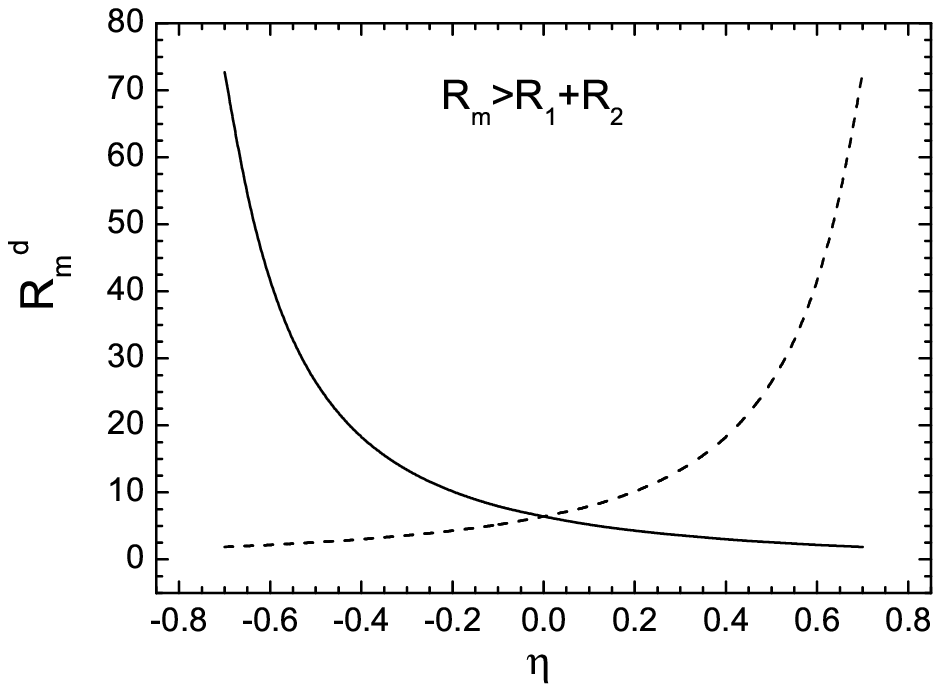}
\includegraphics[width=0.49\linewidth]{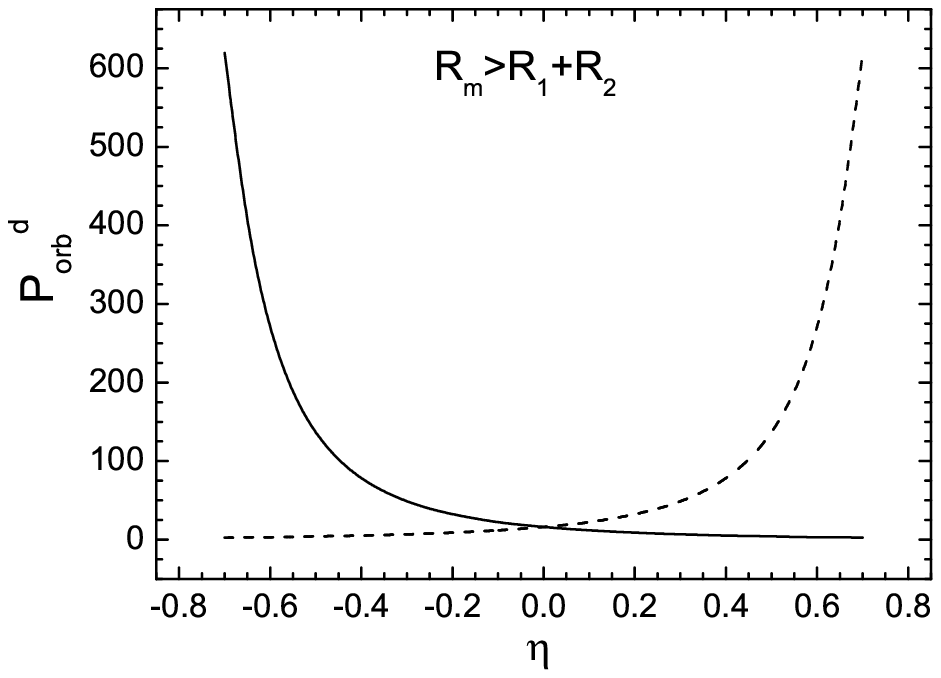}}
\caption{The calculated dimensionless relative distance $R_{m}^{d}$
and orbital rotation period
$ P_{orb}^{d}$ as   functions of mass asymmetry at $R_m>R_t=R_1+R_2$.
 The cases of  antiparallel spins with $L_1=J_{tot}+S_1-S_2$ (dashed line) and
$L_2=J_{tot}-S_1+S_2$ (solid line) are presented.
}
\label{4_fig}
\end{figure}
\begin{figure} [ht]
\centering
{\includegraphics[width=0.5\linewidth]{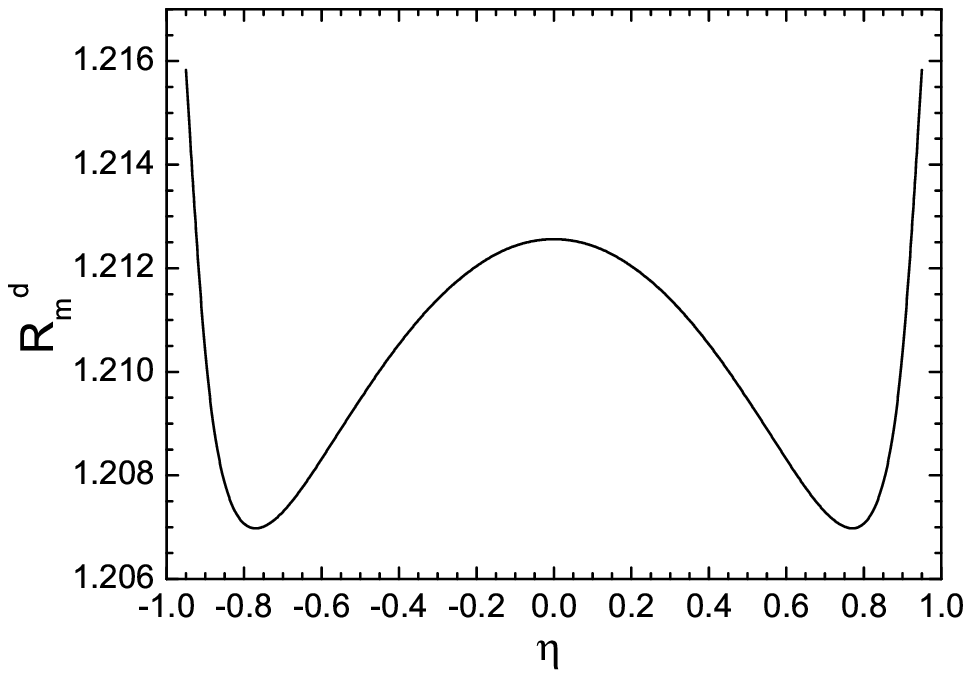}
\includegraphics[width=0.49\linewidth]{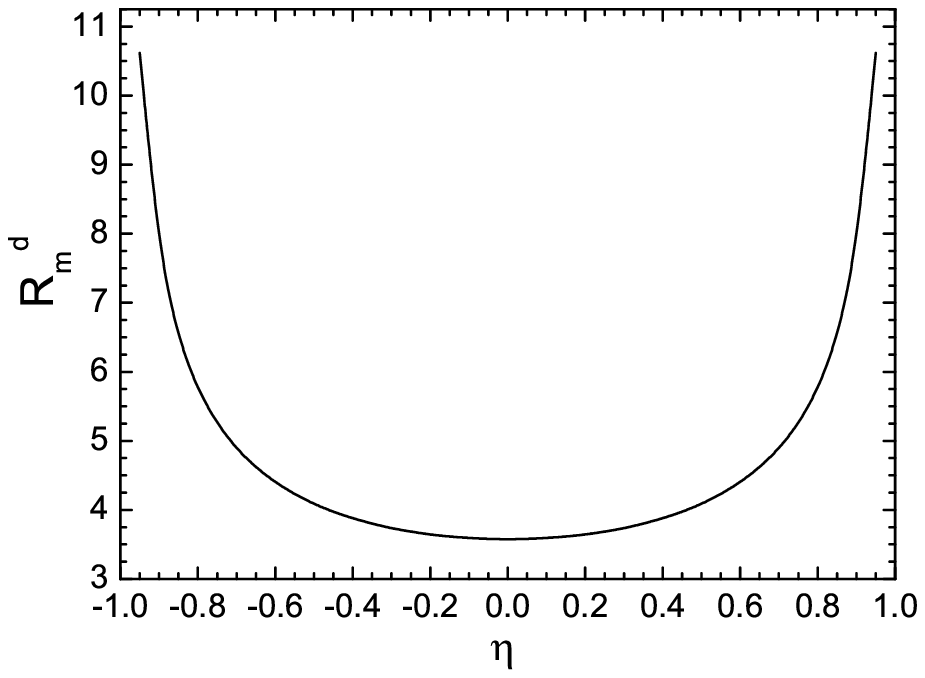}}
{\includegraphics[width=0.49\linewidth]{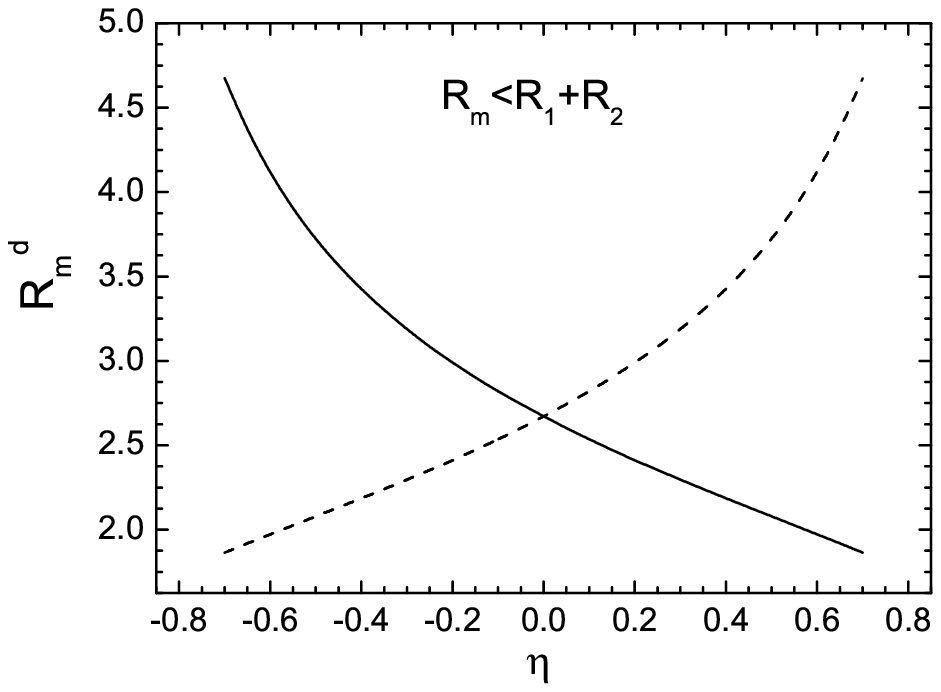}
\includegraphics[width=0.49\linewidth]{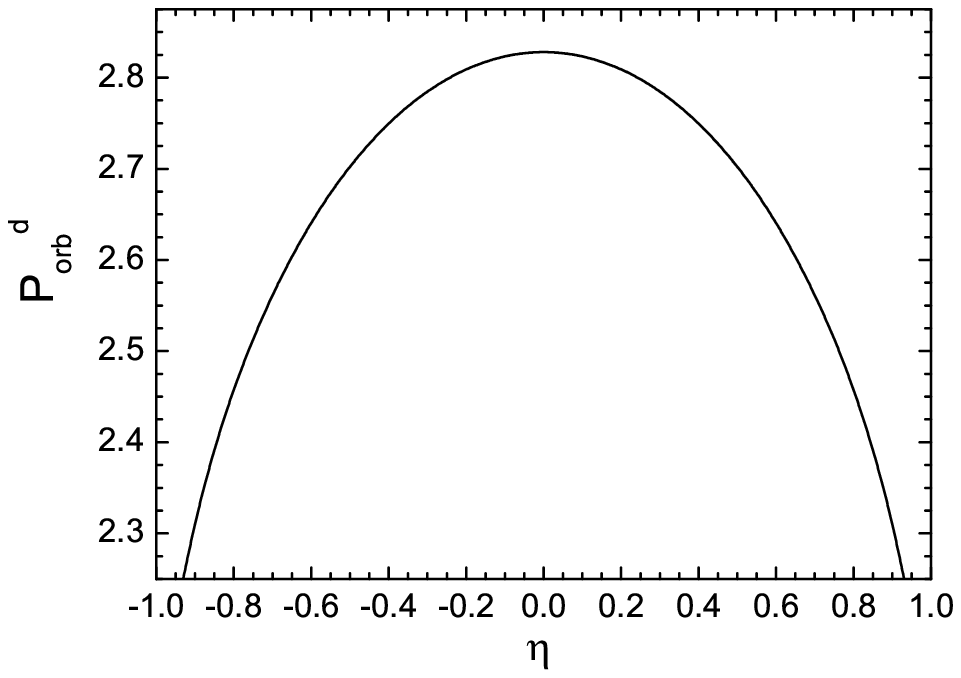}}
\caption{The calculated dimensionless $R_{m}^{d}$ and $ P_{orb}^{d}$
as   functions of mass asymmetry at $R_m\le R_t=R_1+R_2$ and $l=2/3$.
In the first row, the left and right sides correspond to the system with  the  parallel and antiparallel, respectively, orbital and spins
angular momenta.  In the second row, the cases of  antiparallel spins with $L_1=J_{tot}+S_1-S_2$ (dashed line) and
$L_2=J_{tot}-S_1+S_2$ (solid line) are presented.
The period does not depend on the orientations of  orbital and spin
angular momenta.
}
\label{5_fig}
\end{figure}

In conclusion, within   the model \cite{Murad1,Murad2,Murad3} based on the  Regge-like laws,
we have shown that all possible binary stars (planets) or binary galaxies, independently of their mass asymmetry $\eta$,
satisfy the Darwin instability condition ($S_1 + S_2 \ge \frac{1}{3}L$) which contradicts to the observations.
This conclusion is not sensitive to the parameters of model.
Therefore, one should search for other mechanism that triggers
the merger of the contact binary components.

Employing the  Regge-like laws, we have derived the new analytical formulas for the relative distance
and orbital rotation period of the binary system, which depend  on the fundamental constants $G$, $\hbar$, $m_p$,    masses of
the binary components,  and the experimental relation radius-mass. We have predicted
that decreasing and  increasing periods as  functions of mass asymmetry
are related, respectively, with the non-overlapping ($R_m > R_t$)
and overlapping ($R_m \le R_t$) stage of the binary object.

\section{Acknowledgements}

This work was partially supported by  Russian Foundation for Basic Research (Moscow)  and
DFG (Bonn).



\begin{thebibliography}{0}








\bibitem{Cher:2013}
 A.M.  Cherepashchuk, {\it  Close binary stars} (Fizmatlit, Moscow, 2013), vol. I and II.



\bibitem{Kulkarni:2007}  S.R.
Kulkarni, E.O. Ofek, A.   Rau, {\it et al.}, Nature  {\bf 447}, 458 (2007).

\bibitem{Tylenda:2011}
R. Tylenda  {\it et al.},    Astron. Astrophys.
{\bf 528},  A114 (2011).

\bibitem{Soker:2003} N.
Soker and R.  Tylenda,    ApJL  {\bf 582},  L105 (2003).

\bibitem{Molnar:2017}
L.A.  Molnar,  D.M.   Van Noord,  K.   Kinemuchi, J.P.  Smolinski,   C.E.  Alexander,   E.M.   Cook,    B.  Jang, H.A.
Kobulnicky, C.J.  Spedden,    and  S.D. Steenwyk, S.D.  arXiv:1704.05502 (2017).

\bibitem{Abbott:2016a}  B.P. Abbott    {\it et al.}  (LIGO Scientific and Virgo Collaboration), Phys. Rev. Lett.  {\bf 116}, 241102 (2016).

\bibitem{Abbott:2016b}
B.P. Abbott    {\it et al.}  (LIGO Scientific and Virgo Collaboration),    Phys. Rev. Lett. {\bf 116}, 241103 (2016).

\bibitem{Abbott:2017}
B.P. Abbott    {\it et al.}  (LIGO Scientific and Virgo Collaboration),  Phys. Rev. Lett. {\bf  118}, 221101 (2016).


\bibitem{Sana:2012}  H.
Sana,  S.E.  de Mink, A. de Koter,  {\it et al.},   Science  {\bf 337},  444 (2012).

\bibitem{Kocha:2014}  C.S. Kochanek,  S.M.  Adams, and K. Belczynski, MNRAS {\bf 443}, 1319 (2014).

\bibitem{Fabrycky:2007} D.
Fabrycky and S. Tremaine,     Astrophys. J. {\bf 669}, 1298 (2007).

\bibitem{Tokovinin:2006} A.
Tokovinin,  S.  Thomas, M.  Sterzik, and S. Udry,  Astron. Astrophys. {\bf 450}, 681 (2006).


\bibitem{Eggleton:2012} P. P.
Eggleton,  J. of Astronomy and Space Sciences {\bf 29}, 145 (2012).

\bibitem{Pejcha:2017}
O. Pejcha, B.D. Metzger, J.G. Tyles, and K. Tomida, Astrophysical Journal {\bf 850}, 59 (2017).



\bibitem{Darwin:1879} G.H.
Darwin,  Proc. R. Soc. {\bf 29}, 168 (1879).

\bibitem{Rasio:1995} F.A.
Rasio,   ApJL {\bf 444},  L41 (1995).

\bibitem{Stepien:2011} K.
St\c{e}pie\'{n},  Astron. Astrophys. {\bf  531},  A18 (2011).

\bibitem{DSouza:2006}  M.C.R.
D'Souza,  P.M.  Motl, J.E. Tohline, and J. Frank,    Astrophys. J.   {\bf 643}, 381 (2006).

\bibitem{Koenigsberger:2016}G.
Koenigsberger and E. Moreno,  Rev. Mex. Astron. Astrofis. {\bf 52}, 113 (2016).







\bibitem{Murad1} R.M. Muradian,  Astrofiz. {\bf 11}, 237 (1975) [in Russian];
Astrofiz. {\bf 13}, 63 (1977) [in Russian]; Astrofiz. {\bf 14}, 439 (1978) [in Russian].

\bibitem{Murad2} R.M. Muradian,  Astrophys. Space Sci. {\bf 69}, 339 (1980).

\bibitem{Murad3} R.M. Muradian,  Phys. Part. Nucl. {\bf 28}, 471 (1997).

\bibitem{IJMPE} V.V. Sargsyan, H. Lenske, G.G. Adamian, N.V.   Antonenko,
Int. J. Mod. Phys. E {\bf 27},  1850063 (2018); {\bf 27},  1850093 (2018).

\bibitem{nash} V.V. Sargsyan, H. Lenske, G.G. Adamian, N.V.   Antonenko,
Int. J. Mod. Phys. E  (2019) submitted.

\bibitem{Vasil:2012}
 B.V.   Vasiliev,
 Univ. J. Phys. Applic. {\bf 2}, 257 (2014); {\bf 2}, 284 (2014); {\bf 2}, 328 (2014);
 J. Mod. Phys. {\bf 9}, 1906 (2018); {\bf 9}, 2101 (2018).

\bibitem{Karachentsev} I.D. Karachentsev, {\it Binary galaxies}
(Nauka, Moscow, 1987) [in Russian].

\end{thebibliography}
\end{document}